\documentclass[aps,prb,twocolumn,epsf,floatfix]{revtex4}
\usepackage{graphicx}
\preprint{cond-mat/0310293}

\begin{document}

\title{Measurement of two-qubit states by quantum point contacts}
\author{Tetsufumi Tanamoto$^1$ and Xuedong Hu$^2$}
%} 
\affiliation{
$^1$Corporate R\&D Center, Toshiba Corporation,
Saiwai-ku, Kawasaki 212-8582, Japan, \\
$^2$Department of Physics, University at Buffalo, SUNY, Buffalo,
New York 14260-1500,USA  }
%\draft

\begin{abstract}
We solve the master equations of two charged qubits measured by two serially
coupled quantum point contacts (QPCs).  We describe two-qubit dynamics by 
comparing entangled states with product states, and show that the QPC current
can 
be used for reading out results of quantum calculations and providing
evidences of two-qubit entanglement. 
We also calculate the concurrence of the two qubits 
as a function of dephasing rate 
that originates from the measurement.
We conclude that coupled charge qubits can be effectively
detected by a QPC-based detector.
\end{abstract}
\maketitle

\section{Introduction}

Quantum information processing in charge-based solid state nanostructures has
attracted wide spread attention because of the potential scalability of such
devices, and the relative ease 
%
%(compared with systems based on magnetic degrees of freedom such as spin in
%semiconductors and flux in superconductors) 
%
with which such charge devices can be manipulated and
detected.\cite{Nakamura,Makhlin,Fujisawa,Wiel}  Recently, two-qubit coherent
evolution and possibly entanglement have been observed in capacitively
coupled Cooper pair boxes.\cite{Pashkin}  For universal quantum computing,
two-qubit operations are required, so that the realization of controlled
two-qubit entanglement is regarded as a crucial milestone for the study of
solid state quantum computing.  While two-qubit information can be detected
with one measurement device on each qubit, it is also important to search for
a detector that is directly sensitive to two-qubit information, and to
develop a proper formalism to study two-qubit measurement
processes.\cite{Tanamoto,Korotkov,TanaHu}  

The ultimate criterion for the detection of qubits is whether we can
distinguish the results of a quantum computation by the output signal of the
detector, {\it e.g.} current or conductance of a single electron transistor.  
In the case of one qubit, two single-qubit states $|\downarrow\rangle$ and 
$|\uparrow\rangle$ need to be distinguished.  In the case of two qubits, 
four two-qubit state, $|\downarrow \downarrow \rangle$, $|\downarrow
\uparrow \rangle$, $|\uparrow \downarrow \rangle$, and $|\uparrow \uparrow
\rangle$ (we will call them $|A\rangle \sim |D\rangle$) need to be
distinguished before the qubit states are destroyed by the measurement.  As we
mentioned above, measurement of multi-qubit states can generally be achieved
by multiple single-qubit measurements on each of the qubits, respectively. 
Here we study a different detection process: the temporal behavior
of a detector (QPC in the present study) that simultaneously couples to two
qubits.  We show that information contained in the temporal evolution of the
QPC current can help us distinguish different two-qubit product states, and
some entangled states from the product states.  Indeed, one motivation of our
study is to obtain direct evidence for the entanglement of the qubits,
possibly from the detector current or other measurable quantities.
%
%Another motivation of studying direct measurement of two-qubit states
%originate from the measurement of entangled states.  Such measurement can be
%done by first converting an entangled state to a product state using quantum
%operations such as controlled-NOT (CNOT).  If we can then detect the
%resulting product states, the entangled states are detected accordingly. 
%
%$|\uparrow \uparrow \rangle$, ..., $|\downarrow\downarrow \rangle$ 
%($|A \rangle$ $\sim$ $|D\rangle$ in the case of two qubits) by using quantum
%operations such as controlled-NOT (CNOT).  
%However, it is also desirable if we could obtain direct evidence for the
%entanglement of the qubits, possibly from the detector current or other
%measurable quantities.  

%%%%%%%%%%%%% Fig.1
\begin{figure}
\begin{center}
\includegraphics[width=6cm]{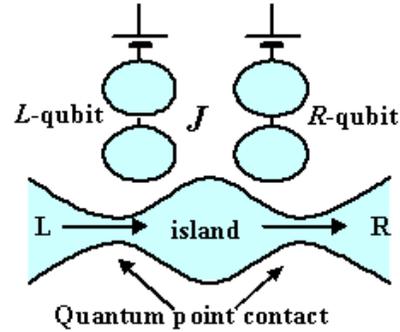}
\end{center}
\caption{Two charge qubits (using double quantum dot charged states) are
capacitively coupled to a detector of two serially coupled QPCs. $J$ is the
strength of inter-qubit interaction.  No tunneling is allowed between the QPC
detector and any of the qubits.}
\label{QPC}
\end{figure}

In Ref.~\onlinecite{TanaHu}, we studied a particular scheme for the quantum
measurement of two charge qubits by a two-island single-electron transistor
(SET), and showed that the SET is an effective detector of the
two-charge-qubit states.  Here the charge qubits are constituted of two
coupled quantum dots (QD) with one excess electron.  Due to tunnel coupling
of the QDs, the wave functions in a qubit are superpositions of localized
states from each of the QDs.  If a qubit is prepared in a single QD state, it
tends to oscillate between the two sides of the double QD.  If we define the
local states as $|\uparrow\rangle$ and $|\downarrow \rangle$, the qubit state
oscillates between the two logical states with a frequency determined 
by the tunneling coupling and the difference of the energy-levels of the two
QDs.  Time-dependent behavior of this coherent oscillation of the qubits is
determined by the initial state.  If we take the initial time to be that 
when a final
quantum operation is applied to the qubit, the detector readout current 
reflects the results of quantum calculation.  The qubit states interact with
the readout current by changing the energy (and therefore occupation) of the
electronic states in the SET islands and possibly the tunneling rates of the
junctions (by modifying the island electronic states themselves) on both
sides of the islands. Although, in Ref.~\onlinecite{TanaHu}, we show that the 
SET can distinguish the different coherent oscillations between the 
two-qubit product states and the entangled states, we have not yet
investigated the two-qubit dynamics itself.

Here we would like to study quantum point contact (QPC) as a detector for two
coupled charge qubits (Fig.~\ref{QPC}).  A QPC is essentially a
one-dimensional conducting channel and is considered to be an effective
charge detector, similar to the SET.  The particular scheme we consider
consists of two low transparency QPCs connected in series through a single
level quantum dot.  Each of the QPCs is close to a charge qubit so that its
current/conductance is dependent on the state of the respective qubit. 
Compared to the SET detector, the QPC detector interacts with qubits only
through the change of tunneling rates.  Although the SET detector is able to
describe a variety of features of the internal states of qubits in
Ref.~\cite{TanaHu}, we could not identify which of the two interactions (that
between the islands and qubits, which modify the level occupations on the SET
islands; or that between the tunneling junctions and qubits, where potential
by the qubits modifies the tunneling rates) plays the major role in the SET
detector.  Thus, an important question is whether or not the QPC detector
that interacts with qubits only by the change of the tunneling rates is also
an effective apparatus for detecting the qubit states.  In this paper,
after discussing the basic two-qubit dynamics with no detector, 
we focus mainly on the following issues: 
(1) whether we can distinguish the four product states $|A\rangle \sim
|D\rangle$ of two coupled charge qubits in the time-domain with a serially
coupled QPC detector, 
(2) whether we can distinguish the entangled states from the product states of
these two qubits, and
(3) whether the quantum Zeno effects in the coupled charge qubits can be
observed.
In the following sections, we solve the master equations
for the coupled qubit-QPC system and investigate the effectiveness of the
proposed QPC detector.
In Sec. II, we present our formulation of two qubits and the QPC detector.
In Sec. III, we discuss the difference between a dynamics of single 
qubit and that of two qubits. In Sec. IV, we show the numerical results 
of two-qubit detection by QPC. Sec. V is devoted to discussion for the 
QPC detection, and Sec. VI consists of a conclusion.

%%%%%%%%%%%%% Fig.2
\begin{figure}
\begin{center}
\includegraphics[width=6cm]{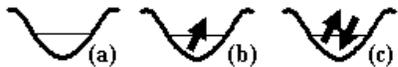}
\end{center}
\caption{Electronic states in the inter-QPC island.  We assume that there is
only one accessible electronic orbital state on the island.  There are thus
totally four possible island states: (a) Empty dot---state $"a"$
has no excess electron on the dot. (b) Single-electron dot---state $"b"$ has
one electron and is spin-degenerate.  (c) Two-electron dot---state $"c"$ has
two electrons in a spin singlet state occupying the same orbital state.}
\label{states}
\end{figure}

%%%%%%%%%%%%%%%%%%%%%%%%%%%%%%%
\section{Formulation}

In the present measurement scheme, the QPCs are capacitively coupled to the
charge qubits (Fig.~\ref{QPC}), so that the current through them sensitively
depends on the states of the qubits.  We describe the two QPCs using two
tunnel matrix elements only and neglect further internal
structures.\cite{Gurvitz2,Aleiner}  
%
%From viewpoint of using the QPCs as a measurement apparatus, this description
%is valid as we assume that the response time of the QPCs are much faster
%than other dynamical times such as the qubit flip time.
%
We assume that the qubit-QPC coupling is purely capacitive, so that there is
no current flowing from the qubits to either of the QPC electrodes.  
%
%This setup is different from those in Josephson charge
%qubits,\cite{Nakamura,Pashkin} where current continues to flow through the
%qubits in the measurement stage.  
%
The Hamiltonian for the combined two qubits and the QPCs is written as $H =
H_{\rm qb}\!+\!H_{\rm qpc}\!+\!H_{\rm int}$.  $H_{\rm qb}$ describes the two
interacting qubits (left and right, as illustrated in Fig.~\ref{QPC}), each
consisting of two tunnel-coupled QDs and one excess charge:\cite{Tanamoto}
%%%%%%%%%%%%%%%%%%%%%%%%%%
\begin{equation}
H_{\rm qb} =\sum_{\alpha=L,R} \left(\Omega_\alpha \sigma_{\alpha x} \!+\!
\Delta_\alpha \sigma_{\alpha z} \right)
\!+\!J \sigma_{Lz} \sigma_{Rz}, 
\end{equation}
%%%%%%%%%%%%%%%%%%%%%%%%%%
where $\Omega_\alpha$ and $\Delta_\alpha$ are the inter-QD tunnel coupling and
energy difference (gate bias) within each qubit.  Here we use the spin
notation such that $\sigma_{\alpha x} \equiv a_\alpha^\dagger b_\alpha +
b_\alpha^\dagger a_\alpha$ and $\sigma_{\alpha z} \equiv a_\alpha^\dagger
a_\alpha - b_\alpha^\dagger b_\alpha$ ($\alpha=L,R$), where $a_\alpha$ and
$b_\alpha$ are the annihilation operators of an electron in the upper and
lower QDs of each qubit.  Thus, $|\!\uparrow\rangle$ and
$|\!\downarrow\rangle$ refer to the two single-qubit states in which the
excess charge is localized in the upper and lower dot, respectively. $J$ is a
coupling constant between the two qubits, originating from capacitive
couplings in the QD system.\cite{Tanamoto}

The two serially coupled QPCs are described by $H_{\rm qpc}$:
%%%%%%%%%%%%%%%%%%%%%%%%%%
\begin{eqnarray}
H_{\rm qpc} \!\!&=&\!\! \!\! 
\sum_{\alpha=L,R \atop s\!=\!\uparrow,\downarrow}\! 
\sum_{\ i_\alpha} \!\!\left[ E_{i_\alpha} c_{i_\alpha s}^\dagger
c_{i_\alpha s} \!+\! V_{i_\alpha s} (c_{i_\alpha s}^\dagger d_{s} +
d_{s}^\dagger c_{i_\alpha s} )\right]
\nonumber \\
&+& \sum_{s\!=\!\uparrow,\downarrow} E_d d_{s}^\dagger d_{s} 
+ U d_{\uparrow}^\dagger d_{\uparrow} d_{\downarrow}^\dagger d_{\downarrow}\,.
\label{eqn:H_qpc}
\end{eqnarray}
\normalsize
%%%%%%%%%%%%%%%%%%%%%%%%%%%
Here $c_{i_{L}s}$($c_{i_{R}s}$) is the annihilation operator of an electron
in the $i_L$th ($i_R$th) level ($i_L(i_R)=1,...,n)$ of the left (right)
electrode, $d_{s}$ is the electron annihilation operator of the island
between the QPCs, $E_{i_{L}s}$($E_{i_{R}s}$) is the energy level of electrons 
in the left (right) electrode, and $E_d$ is that of the island.  Here we
assume only one energy level on the island between the two QPCs, with spin
degeneracy.  $V_{i_Ls}$ ($V_{i_Rs}$) is the tunneling strength of electrons
between the left (right) electrode state $i_Ls$ ($i_Rs$) and the island state.  
$U$ is the on-site Coulomb energy of double occupancy in the island. 
Note that the number of island states here (Fig.~\ref{states}) is much
smaller than that of the two-island SET states,\cite{TanaHu} where we need to
take into account at least 10 types of detector states.  In
Ref.~\onlinecite{TanaHu}, we observed that the two-island SET can represent a
variety of qubit states because of its large number of degrees of freedom. 
With a much simpler state structure for the present coupled QPC scheme, we
will study whether the QPC current could still represent the qubit states
with sufficient clarity.

The capacitive interaction between the qubits and the QPCs is represented by
$H_{\rm int}$, which contains the information that localized charge near the
QPCs increases the energy of the system electrostatically, thus affecting the
tunnel coupling between the QPCs and the island in between:\cite{Gurvitz2} 
%%%%%%%%%%%%%%%%%%%%%%%%%%
\begin{equation}
H_{\rm int} = \sum_{\alpha=L,R} \sum_{i_\alpha,s} \delta V_{i_\alpha s}
( c_{i_\alpha s}^\dagger d_s + d_s^\dagger c_{i_\alpha s} ) \sigma_{\alpha z}
\,,
\end{equation}
%%%%%%%%%%%%%%%%%%%%%%%%%%
where $\delta V_{i_\alpha s}$ is an effective change of the tunneling strength
between the electrodes and QPC island (we shift the origin of the interaction
energy such that $\delta V_{i_\alpha,s} = 0$ corresponds to the case where
qubits are not polarized $\sigma_{\alpha z}=0$).  We assume that the
tunneling strength of electrons weakly depends on the energy $V_{i_\alpha s}
= V_\alpha(E_{i_\alpha s})$ and electrodes are degenerate up to the Fermi
surface.  Then the qubit-QPC interaction dictates that qubit states influence
the QPC tunneling rate in the form of $\Gamma_{\alpha}^{(\pm)} (E) \equiv
2\pi \wp_{\alpha} (E) |V_{\alpha}^{(\pm)} (E) |^2$ and
$\Gamma_{\alpha}^{(\pm)'} (E) \equiv 2\pi \wp_{\alpha} (E\!+\!U) 
|V_{\alpha}^{(\pm)} (E\!+\!U)|^2$, where $V_\alpha^{(\pm)}(E) \!=\!
V_\alpha\!(E) \pm\! \delta V_\alpha$(E) $(\delta V_\alpha(E)>0)$, and
$\wp_{\alpha} (E)$ is the density of states of the electrodes
($\alpha\!=\!L,R$).  
Hereafter, we use $\Gamma_{\alpha}^{(\pm)}$s and $\Gamma_{\alpha}^{(\pm)'}$s
estimated at the Fermi energy $\mu_\alpha$ of the electrodes to describe the
tunneling rate in the detection process of the qubit states by the two QPCs.  
This is reasonable from a practical standpoint since many experiments are
described using $\Gamma_{\alpha}$ \cite{Nakamura,Fujisawa}.  The values of
the corresponding $\Gamma_{\alpha}^{(\pm)}$s are determined by the structure
of the system such as the distance between the qubits and the QPCs.
For example, a $|\!\downarrow\rangle$ state ($|\!\uparrow\rangle$ state) in a
qubit means the excess charge is localized in the lower (upper) dot, so that
the corresponding tunneling rate should be $\Gamma_{L}^{(\!-\!)}$
($\Gamma_{L}^{(\!+\!)}$).  Therefore, two-qubit state $|A \rangle$ would lead
to QPC tunneling rates of $\left( \Gamma_{L}^{(\!-\!)}, \ \Gamma_{R}^{(\!-\!)}
\right)$ or $\left(\Gamma_{L}^{(\!-\!)'}, \ \Gamma_{R}^{(\!-\!)'}\right)$,
depending on whether or not the island between the QPCs is doubly occupied. 
Similarly, $|B \rangle$, $|C \rangle$ and $|D \rangle$ states correspond to
$\left(\Gamma_{L}^{(\!-\!)}, \ \Gamma_{R}^{(\!+\!)}\right)$ [or
$\left(\Gamma_{L}^{(\!-\!)'}, \ \Gamma_{R}^{(\!+\!)'}\right)$],
$\left(\Gamma_{L}^{(\!+\!)}, \ \Gamma_{R}^{(\!-\!)}\right)$ 
[or $\left(\Gamma_{L}^{(\!+\!)'}, \ \Gamma_{R}^{(\!-\!)'}\right)$], and 
$\left(\Gamma_{L}^{(\!+\!)}, \ \Gamma_{R}^{(\!+\!)}\right)$ [or
$\left(\Gamma_{L}^{(\!+\!)'}, \ \Gamma_{R}^{(\!+\!)'}\right)$], respectively.

We construct the equations for the qubit-QPC density matrix elements at zero
temperature $T\!=\!0$, following the procedure developed by
Gurvitz.\cite{Gurvitz2,Gurvitz}  This method is applicable when the energy
level of the inter-QPC island is in between the chemical potentials of the
two electrodes.  The wave function $|\Psi(t)\rangle$ of the qubit-QPC system
can be expanded over the Hilbert space spanned by the two-electron states of
the qubits, the island states of the QPC shown in Fig.~\ref{QPC}, and all
possible electrode states.  We choose $|0\rangle$ to refer to the initial
ground state of the whole detector system (two electrodes and the inter-QPC
island) where the two electrodes are filled with electrons up to $\mu_L$ and
$\mu_R$, respectively, and the two QPCs and the inter-QPC island are empty of
excess electrons.  The basis states for the QPC can then be constructed from
$|0\rangle$ by moving electrons from the left electrode (with higher chemical
potential) to the inter-QPC island and the right electrode.  We categorize
the detector states by the number of electrons that are transferred from the
left to the right electrode (Fig.~1): 
\begin{equation}
|\Psi (t) \rangle \!=|\Psi_0 (t) \rangle \!+\!|\Psi_1 (t) \rangle, 
\label{eqn:wave_all}
\end{equation}
where $|\Psi_0 (t) \rangle$ is the part of the wave function that no electron 
tunnels through to the right electrode and $|\Psi_1 (t) \rangle$ represents
the part of the wave function where one or more electrons are transferred to
the right electrode.  $|\Psi_0 (t) \rangle$ can be expressed as
%%%%%%%%%%%%%%%%%%%%%%%%%%%%%%%%%%%%%%%%%%%%%%%
%\small
\begin{widetext}
\begin{equation}
|\Psi_0 (t) \rangle = \sum_{\small z=A,B,C,D}
\left\{ \ b^{(0)a,z}(t) + 
\sum_{ls} b_{ls}^{(0)b,z}(t) \ d_{s}^\dagger c_{ls} + \sum_{l_1l_2}
b_{l_1l_2\uparrow\downarrow}^{(0)c,z}(t) \ d_{\uparrow}^\dagger
d_{\downarrow}^\dagger c_{l_1\uparrow} c_{l_2\downarrow} \right\} |0 \rangle
|z \rangle \,,
\label{eqn:wave0}
\end{equation}
\normalsize
%%%%%%%%%%%%%%%%%%%%%%%%%%%%%%%%%%%%%%%%%%%%%%%
where $b^{(0)a,z}(t)$, $b_{ls}^{(0)b,z}(t)$ and $b_{l_1l_2 \uparrow
\downarrow}^{(0)c,z}(t)$ are coefficients for the respective states.  The
superscripts refer to the number of electrons transferred ($0$ here), the
states of the QPC island (as illustrated in Fig.~\ref{states}), and the four
two-qubit basis states.  The subscripts refer to the left electrode states
from which electrons tunnel into the islands.  Thus each of the terms in
$|\Psi_0(t) \rangle$ indicates a state with as little as zero but up to 2
electrons moved from the left electrode to the QPC island, while no electron
is transferred to the right electrode.  $|\Psi_1 (t) \rangle$ can be
expressed as
%%%%%%%%%%%%%%%%%%%%%%%%%%%%%%
%\small
\begin{equation}
|\Psi_1 (t) \rangle = \sum_{n=1}^{\infty}
\sum_{\small z=A,..,D \atop \beta_1..\beta_n}
\left\{ b_{\beta_1..\beta_n}^{(n)a,z} (t) + \sum_{l s}
b_{ls\beta_1..\beta_n}^{(n)b,z} (t) \ d_{s}^\dagger c_{l s}
\!+\!\sum_{l_1l_2} b_{l_1l_2\uparrow\downarrow\beta_1..\beta_n}^{(n)c,z}(t) \
d_{\uparrow}^\dagger d_{\downarrow}^\dagger c_{l_1 \uparrow} c_{l_2
\downarrow} \right\} 
\otimes \prod_{i=1}^{n} \left(c_{l_i^\prime s_i^\prime}^\dagger c_{r_i^\prime
s_i^\prime} \right)|0 \rangle |z\rangle \,,
\label{eqn:wave}
\end{equation}
\end{widetext}
\normalsize
%%%%%%%%%%%%%%%%%%%%%%%%%%%%%%%
where $\beta_i \equiv (l_i^\prime,r_i^\prime,s_i^\prime)$ represent the
left electrode, right electrode, and spin states involved in the transferred
electrons.  Similar to the expressions of the coefficients for $|\Psi_0 (t)
\rangle$, here $b_{\beta_1..\beta_n}^{(n)a,z} (t)$, $b_{ls
\beta_1..\beta_n}^{(n)b,z}(t)$ and $b_{l_1 l_2 \uparrow \downarrow
\beta_1..\beta_n}^{(n)c,z}(t)$ are coefficients for the states with $n$
electrons transferred to the right electrode, and another 0 to 2 electrons
moved from the left electrode to the QPC island.  The superscripts again
refer to the number of transferred electrons ($n$), the QPC island states,
and the qubit basis states.  

Substituting this wave function into the Schr\"{o}dinger equation for the
whole qubit-QPC system, $i|\dot{\Psi}(t) \rangle = H|\Psi(t)\rangle$ (having
set $\hbar = 1$), we obtain a set of algebraic equations for the coefficients
in Eq. (\ref{eqn:wave0}) and Eq. (\ref{eqn:wave}).
Assuming zero magnetic field and spin-independent electron tunneling, the
density matrix elements can be defined as
\begin{equation} 
\rho_{u_1u_2}^{z_1z_2}(t) \equiv \sum_{\beta} \int \frac{dEdE'}{4\pi^2} \ 
\tilde{b}_{\beta}^{u_1,z_1}(E) \tilde{b}_{\beta}^{u_2,z_2*}(E) \ e^{i(E - E')
t} \,,
\label{eqn:density}
\end{equation}
where $\tilde{b}_{\beta}^{u_1,z_1}(E)$ is a Laplace-transformed element of 
$b_{\beta}^{u_1,z_1}(t)$ after summing over $\beta=\{0,\beta_1,\beta_2,
\cdots, \beta_n, \cdots\}$, the electrode states of transferred electrons as
discussed above ('0' corresponds to coefficients in Eq.(\ref{eqn:wave0})). 

After a lengthy calculation, we obtain 48 equations for the density matrix
elements $\rho_{u}^{z_1z_2}(t)$, where $u=a,b_\uparrow,b_\downarrow,c$
indicate quantum states of the inter-QPC island (Fig.~\ref{states}) as shown
in Appendix A.\cite{footnote}. Because we assume that there is no magnetic 
field, $\rho_{b\uparrow}^{z_1z_2}(t)=\rho_{b\downarrow}^{z_1z_2}(t)$.

The readout current is obtained as a time derivative of the number of
electrons in the island:\cite{Gurvitz2,Gurvitz,TanaHu}
%%%%%%%%%%%%%%%%%%%%%%%%%%
\begin{equation}
I(t) = e\Gamma_R \left[ \rho_{b\uparrow}(t) + \rho_{b\downarrow} (t) \right] +
2 e\Gamma'_R \rho_c(t) \,,
\end{equation}
%%%%%%%%%%%%%%%%%%%%%%%%%%
where $\rho_{u}$ given by $\rho_{u}\!\equiv\! \rho^{AA}_{u} \!\!+\!
\rho^{BB}_{u} \!\!+\! \rho^{CC}_{u}\!\!+\!\rho^{DD}_{u}$ is the occupation
probability of the particular island state $u$.

As the difference between $\Gamma^{(+)}_\alpha$ and $\Gamma^{(-)}_\alpha$ 
increases, the current difference that depends on the difference of qubit
states increases as well, while the qubits lose their coherence faster due to
the fluctuations in the QPC current, which lead to fluctuations in the qubit
energy levels and thus dephasing.  We quantify the strength of the measurement
by dephasing rates defined as
%%%%%%%%%%%%%%%%%%%%%%%%%%%
\begin{eqnarray}
\Gamma_d^{\alpha} & \equiv & \left( \sqrt{\Gamma^{(+)}_\alpha}
\!-\!\sqrt{\Gamma^{(-)}_\alpha} \right)^2, \nonumber \\
\Gamma_d^{\alpha'} & \equiv & \left( \sqrt{\Gamma^{(+)'}_\alpha}
\!-\!\sqrt{\Gamma^{(-)'}_\alpha} \right)^2 \,,
\label{dephasing}
\end{eqnarray} 
%%%%%%%%%%%%%%%%%%%%%%%%%%%
where $\alpha=L,R$.  These rates are the coefficients of the off-diagonal
density-matrix elements of the time-dependent reduced density matrix
equations for the qubits.  The reduced density matrix elements are
%%%%%%%%%%%%%%%%%%%%%%%%%%%
\begin{equation}
\rho^{z_1z_2} \!\equiv \! \rho^{z_1z_2}_a \!+\! \rho^{z_1z_2}_{b\uparrow}
\!+\! \rho^{z_1z_2}_{b\downarrow} \!+\! \rho^{z_1z_2}_{c}.
\end{equation}
%%%%%%%%%%%%%%%%%%%%%%%%%%%($z_1,z_2=A,B,C,D$).  
This definition of dephasing rate is originally introduced by
Gurvitz\cite{Gurvitz2} for the case of a single qubit.  The dephasing time is
taken as coinciding with the measurement time.  Compared with
Ref.~\onlinecite{Gurvitz2}, where there is a single off-diagonal
density-matrix element, we cannot define a single dephasing rate because of
the complexity of our density-matrix equations.  

Current of a QPC in the tunneling regime is very sensitive to the potential
variations of the QDs that are set close to the QPC channel.\cite{Field}  We
thus can choose from a wide range of parameters for our QPCs.  Here we use a
particular set of representative parameters:
%%%%%%%%%%%%%%%%%%%%%%%%%%%
\begin{eqnarray}
\Gamma_A^L\! & = & \!\Gamma_B^L\! = \Gamma_A^R\! = \!\Gamma_C^R\! =
\!\Gamma^{(-)}\! = \!0.8\Gamma, \\ 
\Gamma_C^L\! & = & \!\Gamma_D^L\! = \Gamma_B^R\!\! = \!\Gamma_D^R\!\! =
\!\Gamma^{(+)}\! = \!1.2\Gamma, 
\end{eqnarray}
%%%%%%%%%%%%%%%%%%%%%%%%%%%
which lead to $\Gamma_d^L = \Gamma_d^R (\equiv \Gamma_d) \sim 0.04\Gamma$ for
a typical case ($\Gamma$ is a tunneling rate in the absence of the qubits, so
that dephasing rate is more than one order of magnitude smaller,
corresponding to a relatively weak measurement).  
We can regard $\Gamma_d^{-1}$ as the typical measurement time.  
Obviously, the qubit dynamics (represented by
tunneling rate $\Omega$) would be able to be detected 
when $\Omega^{-1}<\Gamma_d^{-1}$. 
Because, in the present setup, the current without qubits saturates in 
the range of $\sim \Gamma^{-1}$, the time $\Gamma^{-1}$ would 
serve as a standard of when measurement started.
% Obviously, the qubit dynamics (represented by
% tunneling rate $\Omega$) that we should be able to observe with such a QPC
% should fall in the range of $\sim \Gamma$ and $\Gamma_d$: the QPC response
% would not be able to track something much faster than $\Gamma$, while
% anything that develops much slower than $\Gamma_d$ would be washed out by
% the measurement.  
We can also incorporate the effect of Coulomb interaction by
setting $\Gamma_L^{(\pm)'} = 0$ as a limiting case of strong on-site Coulomb
blockade ($U \rightarrow \infty$ in Eq.~(\ref{eqn:H_qpc}) so that no double
occupation is possible), while for weak Coulomb interaction on the island we
can set $\Gamma_\alpha^{(\pm)'} = \Gamma_\alpha^{(\pm)}$ at the limit of
$U=0$.

%%%%%%%%%%%%%%%%%%%%%%%%%%%%%%%%%%%
\section{Qubit dynamics without detector}
%%%%%%%%%%%%%%%%%%%%%%%%%%%%%%%%%%%
In order to better understand our numerical results and the capability of our
QPC detector, it is instructive to first examine the dynamics of both a
single qubit and two qubits in the absence of any detector, and discuss how
the qubit dynamics is measured by the detector.  

We first solve the density matrix equations for a single
qubit on the basis of localized single quantum dot states 
$|\uparrow\rangle$ and $|\downarrow\rangle$:
%------[new@@0409]s
\begin{eqnarray}
\dot{\rho_{\uparrow\uparrow}}&=&
i\Omega(\rho_{\uparrow\downarrow}-\rho_{\downarrow\uparrow}), 
\label{eqn:dm_pr1}\\
\dot{\rho_{\downarrow\downarrow}}&=&
i\Omega(\rho_{\downarrow\uparrow}-\rho_{\uparrow\downarrow}), 
\label{eqn:dm_pr2}\\
\dot{\rho_{\uparrow\downarrow}}&=&
i\Delta \rho_{\uparrow\downarrow}+
i\Omega(\rho_{\uparrow\uparrow}-\rho_{\downarrow\downarrow}).
\label{eqn:dm_pr3}
\end{eqnarray}
For the simple case of $\Delta=0$ (no voltage bias between the two dots so
that qubit dynamics is completely determined by the inter-dot tunnel coupling
$\Omega$, which corresponds to the optimal operational point 
in terms of minimum dephasing as discussed in Ref.\cite{Vion}), 
and starting from one of the localized states $\uparrow$-state
($\rho_{\uparrow\uparrow} (t=0) = 1$) or $\downarrow$-state
($\rho_{\downarrow\downarrow}(t=0)=1$), 
we have
%the solutions for the density matrix elements are
%%%%%%%%%%%%%%%
\begin{eqnarray}
\rho_{\uparrow\uparrow}(t) &=&
\rho_{\uparrow\uparrow}(0) \cos^2(\Omega t) +
\rho_{\downarrow\downarrow}(0) \sin^2(\Omega t), \\
\rho_{\downarrow\downarrow}(t) &=&
\rho_{\downarrow\downarrow}(0) \cos^2(\Omega t) +
\rho_{\uparrow\uparrow}(0) \sin^2(\Omega t), \\
\rho_{\uparrow\downarrow}(t) &=&
\frac{i}{2}(\rho_{\uparrow\uparrow}(0)
-\rho_{\downarrow\downarrow}(0)) \sin(2\Omega t). 
\end{eqnarray}
These solutions dictate that the QPC current should essentially be determined
by $\rho_{\uparrow\uparrow} (t) - \rho_{\downarrow\downarrow} (t) =
[\rho_{\uparrow\uparrow} (0) - \rho_{\downarrow\downarrow} (0)] \cos 2\Omega
t$.  The oscillatory component of the QPC current should thus be dominated by
a $2\Omega$ component (in the case of $\Delta\neq 0$, $2 \sqrt{\Omega^2 +
\Delta^2/4}$), and the temporal evolution of the current is intimately related
to the initial state.

We can also infer information on the two-qubit product states from the
detector current in a similar manner because density matrices of the product
states are written as  $\rho^{AA}(t) = \rho_{\downarrow \downarrow}^L (t)
\rho_{\downarrow \downarrow}^R (t)$ and so on.  Here we solve the two-qubit
dynamics in the absence of the detector by expanding the density matrix on
the basis spanned by the Bell states: $|e_1\rangle \!\!=\!\!
(|\!\!\downarrow \downarrow \rangle \!+\!|\!\uparrow \uparrow
\rangle)/\sqrt{2}$, $|e_2\rangle \!\!=\!\! (|\!\!\downarrow \downarrow
\rangle \!-\!|\!\uparrow \uparrow \rangle)/\sqrt{2}$, $|e_3\rangle \!\!=\!\!
(|\!\!\downarrow \uparrow \rangle \!+\!|\!\uparrow \downarrow
\rangle)/\sqrt{2}$, and $|e_4\rangle \!\!=\!\! (|\!\!\downarrow \uparrow
\rangle \!-\!|\!\uparrow \downarrow \rangle)/\sqrt{2}$ (singlet state). 
%
% A study of the time evolution of two charge qubits in the Bell basis (in the
% absence of the QPC detectors) can help us better understand the features of
% Figs.~\ref{entangled} and \ref{general}.  
%
If we assume two identical qubits ($\Omega_L=\Omega_R$ and $\Delta_L =
\Delta_R (=\Delta)$), the density matrix equations for the two qubits
(without the QPC detector: $\Gamma_d^\alpha=0$) are written as
%%%%%%%%%%%%%%%%%%%%%%%%%%%%%%%%%%%%
\begin{eqnarray}
& & \left\{\begin{array}{ll}
\dot{\rho}^{e_4e_4}&=0 \\
\dot{\rho}^{e_2e_2}&=2i\Delta (\rho^{e_2e_1}-\rho^{e_3e_2})  \\
\dot{\rho}^{e_2e_4}&=-2iJ\rho^{e_2e_4}-2i\Delta\rho^{e_1e_4}
\end{array}
\right. \left. \right.\label{eqn:e1}\\
& &  \left\{\begin{array}{ll}
\dot{\rho}^{e_1e_1}&= 2i\Omega (\rho^{e_1e_3}-\rho^{e_3e_1})
\!+\!2i\Delta(\rho^{e_1e_2}-\rho^{e_2e_1})  \\
\dot{\rho}^{e_3e_3}&=-2i\Omega (\rho^{e_1e_3}-\rho^{e_3e_1})  \\
\dot{\rho}^{e_1e_3}&=-2i\Omega (\rho^{e_3e_3}-\rho^{e_1e_1}) 
\!-\!2i J \rho^{e_1e_3}\!-\!2i\Delta \rho_{e_2e_3}
\end{array}
\right. \label{eqn:e2}\\
& & \left\{\begin{array}{ll}
\dot{\rho}^{e_1e_2}&=-2i\Omega \rho^{e_3e_2}
\!-\!2i\Delta (\rho^{e_1e_1}\!-\!\rho^{e_2e_2}) \\
\dot{\rho}^{e_2e_3}&= 2i\Omega \rho^{e_2e_1}
\!-\!2i J \rho^{e_2e_3}\!-\!2i\Delta \rho^{e_1e_3} 
\end{array}
\right.  \label{eqn:e3} \\
& & \left\{\begin{array}{ll}
\dot{\rho}^{e_3e_4}&=-2i\Omega \rho^{e_1e_4} 
\\
\dot{\rho}^{e_1e_4}&=-2i\Omega \rho^{e_3e_4} 
\!-\!2iJ \rho^{e_1e_4}\!-\!2i\Delta \rho^{e_1e_4}
\end{array}
\right. \label{eqn:e4}
\end{eqnarray}
%%%%%%%%%%%%%%%%%%%%%%%%%%%%%%%%%%%%%
If $\Delta=0$, which again corresponds to the optimal operational point,
% and is the case we will discuss here, 
the density matrix equations can be classified
into four groups (indicated by the four parentheses above).  First of all,
Eqs.~(\ref{eqn:e1}) shows that the probabilities in $|e_2 \rangle$ and $|e_4
\rangle$ states are time-independent.  On the other hand, according to
Eq.~(\ref{eqn:e2}), the probabilities in $|e_1\rangle$ and $|e_3\rangle$
states oscillate as a function of $\{\cos (4\Omega^* t), \sin (4\Omega^*
t)\}$ ($\Omega^*\equiv \sqrt{\Omega^2 \!+\!J^2/4}$).  Meanwhile, 
Eqs.~(\ref{eqn:e3}) and (\ref{eqn:e4}) indicate that the off-diagonal
elements $\{\rho^{e_1e_2}, \rho^{e_2e_3}, \rho^{e_3e_4}, \rho^{e_1e_4}\}$ 
contain $\{\cos (2\Omega^* t), \sin (2\Omega^* t)\}$ type of oscillations.
Therefore, the occupation probabilities for the product states,  
$\rho^{AA}\!\!=\!\! (\rho^{e_1e_1}\!+\!\rho^{e_1e_2}\!+\!
\rho^{e_1e_2}\!+\!\rho^{e_2e_2})/2$, 
$\rho^{BB}\!\!=\!\! (\rho^{e_3e_3}\!+\!\rho^{e_3e_4}\!+\!
\rho^{e_3e_4}\!+\!\rho^{e_4e_4})/2$, 
$\rho^{CC} \!\!=\!\! (\rho^{e_3e_3}\!-\!\rho^{e_3e_4}\!-\!
\rho^{e_3e_4}\!+\!\rho^{e_4e_4})/2$, and
$\rho^{DD} \!\!=\!\! (\rho^{e_1e_1}\!-\!\rho^{e_1e_2}\!-\!
\rho^{e_1e_2}\!+\!\rho^{e_2e_2})/2$,
should all contain $\{\cos (2\Omega^* t), \sin (2\Omega^* t)\}$ oscillations, 
reconfirming the calculations on single-qubit dynamics at the beginning of
this section.  Therefore, we should be able to distinguish pure entangled
states from pure product states $|A\rangle$ $\sim$ $|D\rangle$ based on
whether the detected period of the coherent oscillations is limited to $\{
\cos (4\Omega^* t), \sin (4\Omega^* t)\}$ ($|e_1 \rangle$ and $|e_3\rangle$)
or time-independent ($|e_2 \rangle$ and $|e_4\rangle$) in the limit of weak
interaction between the qubits and the QPCs.  Such behavior is indeed evident
in our results as shown in the following section.

%%%%%%%%%%%%%%%%%%%%%%%%%%%%%%%%%%%
\section{Numerical results of QPC detection}
%%%%%%%%%%%%%%%%%%%%%%%%%%%%%%%%%%%

In Ref.\onlinecite{TanaHu}, we clarified three major issues regarding the
capability of the two-island SET by monitoring its time-dependent readout
current: (1) the two-qubit product states (eigenstates in the absence of
inter-qubit interaction and inter-dot coupling within each qubit) $|A\rangle
\ \sim \ |D\rangle$ can be distinguished; (2) pure entangled states and pure
product states can be distinguished; (3) quantum Zeno effect is present in a
two-qubit system.  In the following we show that similar results are obtained 
for the serially-coupled QPC detector despite its simpler state structure.  

\begin{figure}
\includegraphics[width=8.2cm]{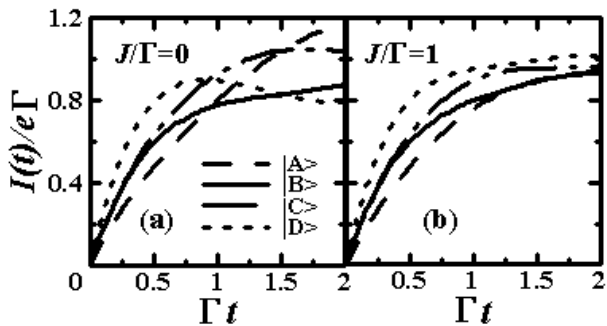}
\caption{Time-dependent QPC current $I(t)$ of the $U=0$ case
($\Gamma_\alpha^{(\pm)'} \!=\! \Gamma_\alpha^{(\pm)}$) starting from four
product qubit states $|A\rangle$ $\sim$ $|D\rangle$ at time $t=0$.  $\Omega_L
= \Omega_R = 0.75\Gamma$, $\Gamma_d = 0.04 \Gamma$.  The two panels refer to
two different inter-qubit interaction: (a) $J=0$, (b) $J=\Gamma$.  We can
distinguish the four product states in both the $J=0$ case and the $J=\Gamma$
case.  This shows that we can distinguish the four two-qubit product states
in a range of inter-qubit coupling strength.}
\label{I-V}
\end{figure}

Figure~\ref{I-V} shows the time-dependent current at small time $t\sim 0$ in
the case of weak Coulomb interaction ($U=0$) ($\Gamma_\alpha^{(\pm)'} \!=\!
\Gamma_\alpha^{(\pm)}$) assuming that initially the two qubits are in one of
the four product states.  To calculate the current when the two-qubit initial
state is $|A\rangle$, for example, we set $b^{(0)a,A}(0) = 1$ and the other
coefficients to zero in the total wave function (Eqs.~(\ref{eqn:wave0}) and
(\ref{eqn:wave})), which means that $\rho_{aa}^{AA}(0) = 1$ and other density
matrix elements are all zero at $t = 0$.  At small $t$ initial state
$|A\rangle$ (with both electrons located in the respective lower dots) leads
to the strongest suppression of the QPC current, while initial state
$|D\rangle$ (with both electrons located in the respective upper dots) the
least.  States $|B\rangle$ and $|C\rangle$ also produce different QPC
currents.  The reason is that there is a finite bias between the left and
right electrodes, so that current flows only in one direction.  Consequently,
$|C\rangle$, with the left qubit electron in the upper dot (thus less
suppression on current), produces a faster rise in current than $|B\rangle$. 
Since the product states are not the two-qubit eigenstates, they evolve into
superposition states and the corresponding QPC current oscillates. 
Nevertheless, we can distinguish the four initial product states by the
values of the readout current in both $J=0$ and $J \neq 0$ cases.  Hereafter
we will focus on the $J=0$ case.  As shown in Fig.~\ref{I-V}, the current
differences between the four two-qubit states can be detected before the
coherent motion of the qubits changes the two-qubit state as $\Omega t
<\pi/4$.  

\begin{figure}
\includegraphics[width=8.2cm]{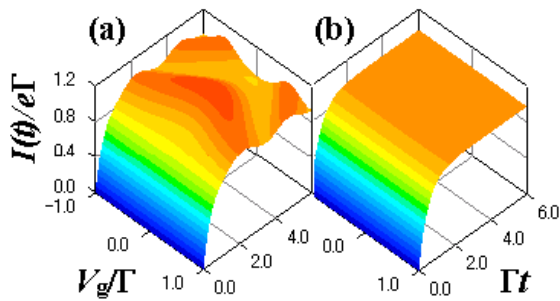}
\caption{Time evolution of QPC current $I(t)$ corresponding to the product
$|B\rangle$ state (panel (a)) and the entangled singlet state $|e_4\rangle$
(panel (b)) when the qubit gate-bias $V_g (=\Delta_L=\Delta_R$) changes.  The
relevant parameters are chosen as $\Omega_L=\Omega_R=0.75\Gamma$, $J=0$,
$U=0$ and $\Gamma_d=0.04\Gamma$.  The $I(t)$ for the product state ((a))
explicitly reflects the coherent oscillations of the qubit states, while 
those for the entangled state are rather uniform.  
%
%In particular, the singlet state is most robust to the measurement.
}
\label{entangled}
\end{figure}

\begin{figure}
\begin{center}
\includegraphics[width=5cm]{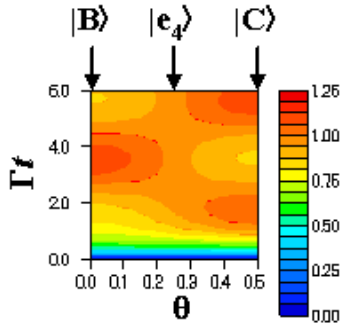}
\caption{A contour plot of the time evolution of QPC current for states
ranging between $|B\rangle$ state and $|C\rangle$ state through singlet state
$|e_4\rangle$ (see text).  The current for the ``general" singlet state shows
uniform characteristics when it is close to the exact singlet state
$|e_4\rangle$ as $|\theta \!\pm\! \pi/4 |\!\stackrel{<}{=}\!\pi/12$.  The
chosen parameters are similar to what we have before: $\Omega_L = \Omega_R =
0.75\Gamma$, $J=0$, $V_g\!=\!0$, $U=0$, and $\Gamma_d=0.04\Gamma$.  In
addition, the current for $|C\rangle$ has a oscillatory component of
frequency $2\Omega=1.5$.
}
\end{center}
\label{general}
\end{figure}

%%%%%%%%%%%%%%%%%%
% entangled state
%%%%%%%%%%%%%%%%%%
%[new@0406]
One observation we made for charge qubits measured by an SET detector is that 
the amplitude of the SET current oscillations corresponding to the pure
entangled states are smaller than those of the pure product
states.\cite{TanaHu}  Similar effects are also observed for the QPC detectors
here, as indicated in Figs.~\ref{entangled} and \ref{general}.
%%%%%%%%%%%%%%%%
% singlet
%%%%%%%%%%%%%%%% Fig.6
A qualitative reason is that the wave functions of the entangled states in the
charge qubits extend over both qubits compared to the product states, so that
the charge distribution of the entangled states is less effective in
influencing the readout current.  Quantitatively, for instance,
Eq.~(\ref{eqn:e1}) also dictates that current corresponding to a singlet
state should have very weak time dependence.  Indeed, Fig.~\ref{entangled}
shows strong differences between QPC currents for the singlet state 
$|e_4\rangle$ and product state $|B\rangle$: the detector current on the
singlet state is essentially a smoothly rising function of time, while the
current for the product state has an oscillatory component of frequency $\sim
2\Omega$ at $V_g=0$.  We obtained similar current behaviors for other
entangled states and product states, where the peaks of the coherent
oscillations in the other product states are shifted as
inferred from Fig.~\ref{I-V}.  These features hold as long as the entangled
states are close to any of the Bell states, 
$|e_1\rangle$, $\sim$ $|e_4\rangle$. 
Figure~\ref{general} shows the time-dependent current of the generalized
singlet state $\cos\theta |\!\downarrow \uparrow \rangle \!+\!e^{i\varphi}
\sin\theta |\!\uparrow\downarrow\nobreak\rangle$ in the range of
$\varphi=\pi$, $0\!\stackrel{<}{=} \!\theta \!\stackrel{<}{=} \!\pi/2$.  We
found that the uniformity of the readout current holds approximately up to
$|\theta \!\pm\! \pi/4 |\!\stackrel{<}{=}\!\pi/12$, which is remarkably
robust (similar to the case of charge qubits measured by SET
detectors\cite{TanaHu}).  
In addition, in Fig.~\ref{general},
the current for $|C\rangle$ state, another product state, also contains an
oscillatory component of frequency $\sim 2\Omega$.
%%%%%%%%%%%%% Fig.3

%%%%%%%%%%%%% Fig.4

\begin{figure}
\includegraphics[width=8.2cm]{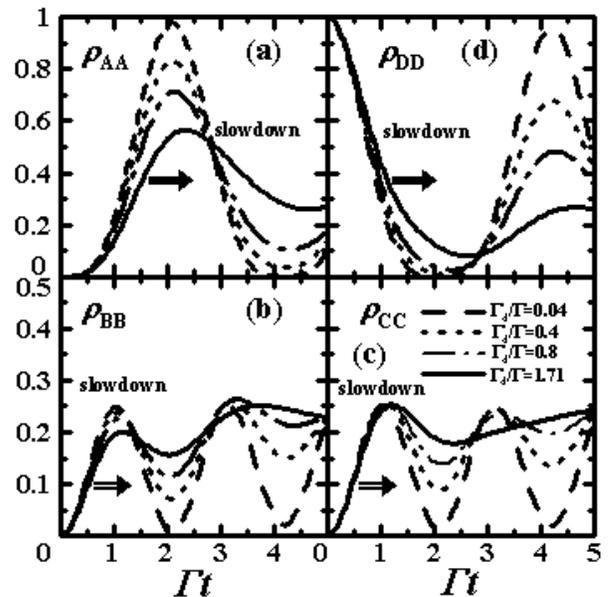}
\caption{Time-dependence of $\rho^{AA}(t)$,$\rho^{BB}(t)$,
$\rho^{CC}(t)$ and $\rho^{DD}(t)$ for the $U=0$ case  ($\Gamma_\alpha^{(\pm)'}
\!=\! \Gamma_\alpha^{(\pm)}$), starting from $\rho^{DD}(t=0)=1$, for
different measurement strengths (in terms of $\Gamma_d$).  Here the
intra-qubit tunneling rates are $\Omega_L=\Omega_R=0.75\Gamma$, and there is
no interaction between the qubits: $J=0$.  As measurement strength $\Gamma_d$
increases, the coherent motions of qubits slow down, which is a clear
evidence of the quantum Zeno effect.}
\label{zeno}
\end{figure}

%%%%%%%%%%%%%%%% Fig 6
An interesting aspect in studying quantum measurement is to explore the
backaction of the measurement apparatus on the qubits.  In this context, the
quantum Zeno effect refers to the phenomenon that a continuous measurement
slows down transitions between qubit quantum states due to the collapse of
the qubit wave function into observed states.  This phenomenon might be useful
in quantum computation because it preserves the results of quantum
calculations for a longer period of time.\cite{Duan}  Figure~\ref{zeno}
demonstrates the quantum Zeno effect for two qubits measured by the QPC, 
where the initial state is chosen to be $|D\rangle$ state ($\rho^{DD}(t=0) =
1$).  As the measurement strength increases ($\Gamma_d$ increases), the
oscillations of the density matrix elements of the two qubits are delayed,
which is clear evidence of the slowdown described by the Zeno effect.

\begin{figure}
\begin{center}
\includegraphics[width=6.0cm]{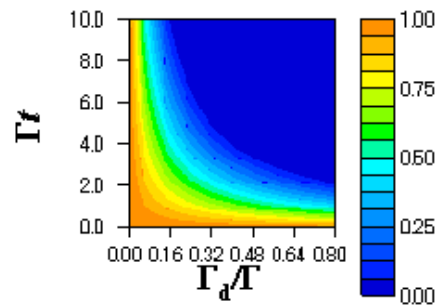}
\caption{Time dependent concurrence of a two-qubit state starting 
from a singlet state $|e_4\rangle$ as a function of the dephasing rate
$\Gamma_d$ in the same parameter region as Fig.\ref{general}.
At $t=0$ and $\Gamma_d=0$, the concurrence takes a value of 1 
and rapidly decreases to zero for large dephasing rates.
}
\end{center}
\label{concurrence}
\end{figure}

\begin{figure}
%\begin{center}
\includegraphics[width=8.2cm]{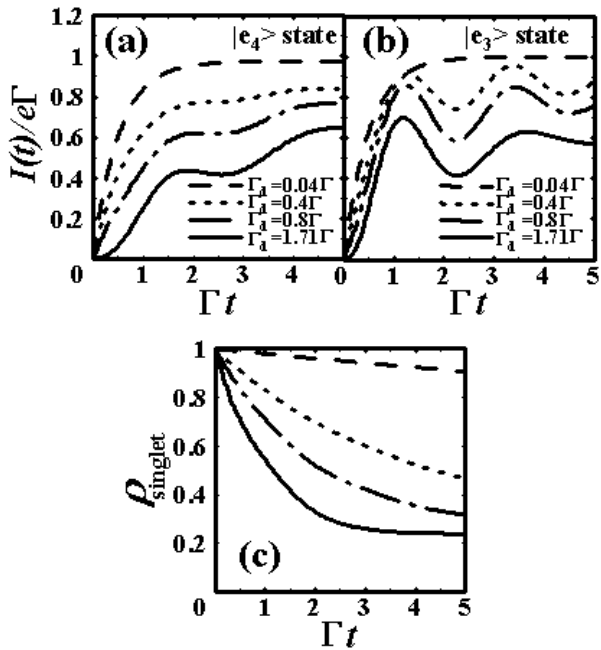}
\caption{Time dependent currents for $|e_4\rangle$ (singlet state) (panel (a))
and $|e_3\rangle$ state (panel (b)),  and the diagonal density matrix
element (panel (c)) for the singlet state, when the dephasing
rate $\Gamma_d$ is changed with $\Delta=0$.  
The parameters are the same as those
in Fig.~\ref{general}.  At $\Gamma_d<0.04\Gamma$, 
Fig.(a) presents the proof of time-independence of the singlet state 
in Eq.(\ref{eqn:e1}), and Fig.(b) shows the proof of 
the $4\Omega$ oscillation of Eq.(\ref{eqn:e2}). 
Figure (c) shows that the two-qubit state begins to
include states other than the singlet states, resulting in the oscillation 
of the current (panel (a)) when $\Gamma_d$ becomes large.}
%\end{center}
\label{new}
\end{figure}

In general, increasing measurement strength (i.e. the coupling strength
between the qubits and the QPCs) leads to faster entanglement between the
qubits and the measuring apparatus, so that measurement leads to projection
of qubit states into product states.  Therefore, stronger measurement strength
destroys entangled qubit states more rapidly.  This is in contrast to the
product states, for which the quantum Zeno effect is observed
(Fig.~\ref{zeno}).\cite{Duan}  We use the concept of
concurrence\cite{Wootters} to quantify two-qubit entanglement and calculate
concurrence in the presence of increasing measurement strength.
%
%Here we estimate the measure of entanglement from the reduced density matrix
%of the qubits using the concept of concurrence.  
%
Figure~\ref{concurrence} shows the effect of measurement on the singlet state,
demonstrating that stronger measurement (in the form of larger $\Gamma_d$)
degrades the entanglement (in terms of concurrence) more rapidly.  
%%%[new@@0409]s
As seen from Eqs.(\ref{eqn:dm_pr1})-(\ref{eqn:dm_pr3}) and from
(\ref{eqn:e1})-(\ref{eqn:e4}), product states and entangled states discussed
here are generally not two-qubit eigenstates even in the absence of the
detector, and thus could evolve into each other through the time-dependent
coherent oscillations.  Strong detection enhances the mixing of these states
and makes it more difficult to infer the qubit states from the detector
current.  
Figures~\ref{new} (a) and (b) show the time-dependent currents 
of $|e_4\rangle$ (singlet state) and $|e_3\rangle$ state
as functions of increasing measurement strength. 
Without the detector, singlet $|e_4\rangle$
state should be time-independent according to Eq.(\ref{eqn:e1}),
and $|e_3\rangle$ should show $4\Omega$ oscillation according 
to Eq.(\ref{eqn:e2}). 
Figures~\ref{new} (a) and (b) indicate these characteristics in the 
weak measurement case $\Gamma_d <0.04\Gamma$, which is also the case 
that we discussed concerning Fig.~\ref{entangled}. 
In this region, we would be able to distinguish the different behaviors 
of entangled states and product states. 
However, as the strength of measurement increases, the detector 
current starts to acquire other oscillatory components, which means that
both states are mixing with other states after $t=0$. 
Figure~\ref{new} (c) is a time-dependent diagonal matrix element 
$\rho^{e_4e_4}$ of the singlet state. This figure also shows that 
the singlet state mixes with other states as the strength of 
measurement increases.

%%%%%%%%%%%%% Fig.5

%%%%%%%%%%%%% Fig.6

%%%%%%%%%%%%% Fig.7

%%%[new@0409]e
In the case of a strong Coulomb interaction so that $\Gamma_L^{(\pm)'}=0$, we
have obtained similar results, except that the magnitude of the average
current is reduced by half because the onsite Coulomb interaction closes one
transmission channel.  This is different from the coupled SET detector we
studied before, where current uniformity in finite-$U$ model is more
persistent than in the infinite-$U$ model, because the internal degrees of
freedom in the two-island SET allow a redistribution of electrons between
the islands.  Here there is only one island with three island states
(unoccupied, singly occupied, and doubly occupied, shown in
Fig.~\ref{states}).  The much simpler internal dynamics of these states is
insufficient to cause any large change in the QPC current when Coulomb
interaction is accounted for. 
%%%%%%%%%%%%% Fig.8 (new)
%%%%%%%%%%%%%%%%%%%%%%%%%%%%%%%%%%%%%%%%
\section{Discussion}
%%%%%%%%%%%%%%%%%%%%%%%%%%%%%%%%%%%%%%%%
In our study so far we have demonstrated that two-charge-qubit state
information can be clearly revealed by the transient current variations in a
serially coupled QPC charge detector.  An important question is then whether
such current evolutions are experimentally observable.  In our calculation, 
$\Gamma$, the QPC tunneling rate, is the physical quantity that can be
directly connected to experiments.  For example, for $\Gamma$ in the order of
100 MHz, Figs.~\ref{I-V}, \ref{entangled} and \ref{new} show that our scheme
requires measurement of a 1 pA current that changes in the nanosecond time
scale.  This is at the edge of the current technology that allows the
measurement of 1 pA current with dynamics in the GHz frequency range with
repeated-measurement
technique.\cite{Nakamura,Fujisawa,Schoelkopf,Gardelis,Cain,Potok}

%[new@0406]start%
One issue we have been trying to address in this study is to compare the
measurement capability of a QPC detector and an SET detector.  In terms of
the theoretical descriptions of the qubit-detector interaction, the major
difference between the QPC detector studied here and the SET detector in
Ref.\cite{TanaHu} is that we model each QPC by a tunnel junction
(Ref.\cite{Gurvitz2}), so that the QPC-qubit interaction directly modifies
strength of tunneling, while in Ref.\cite{TanaHu}, the SET-qubit interaction
influences both the SET island state energy and the island-lead tunneling. 
Despite these differences, our numerical results showed that the current
through the coupled QPC exhibits behaviors similar to those of the 
two-island SET current
in many respects, such as in distinguishing the different qubit product
states, in distinguishing the Bell-type entangled states from the product
states, and in the observation of quantum Zeno effect for the qubit product
states.  
%
% The shared current characteristics seem to indicate the common feature 
% of these two
% types of detectors that they are both capacitively coupled to the charge
% qubits, and the interaction directly changes the tunnel coupling within the
% detectors.  
%[new@0405]end%
%
Stronger differences between QPC and SET detectors do appear when the
qubit-detector interaction strength increases.  The measurement current of the
detector that has a larger number of internal degrees of freedom (the
two-island SET) seems to be able to describe more elaborate quantum states of
the two qubits.  For example, the SET current can clearly distinguish the
four product states shown in Ref.\onlinecite{TanaHu}, while with the present
QPC detector the current shows a simpler structure and smaller differences
for the different qubit states.
%, though the four product states can still be distinguished.  
Qualitatively, the tunneling rate of a QPC is generally
larger than that of an SET, which corresponds to shorter dwelling time for
the QPC (in the present study the dwelling time for QPCs is effectively taken
to be zero).  This difference essentially originates from the simpler
structure of a QPC compared to an SET.
%
%To obtain stronger current differences, the change of the tunneling rate (due
%to charge movement in the qubits) needs to be larger.  This requirement can
%lead to difficulties for a QPC detector in the fabrication stage.  However,
%the simpler structure of QPC compared with the more complicated SET
%structure should at least partially compensate for this problem.  In any
%case, our studies show that both SET and QPC are capable of detecting
%two-charge-qubit states.

In the present study we obtained the density-matrix equations under the
condition that the voltage bias between the left and right electrodes is
sufficiently large such that the left-right symmetry is broken and the
transmission of electrons from the right electrode to the left can be
neglected.  Thus we cannot directly calculate the QPC differential
conductance, which would provide more information for some
experiments.\cite{Schoelkopf}  This is one of the limitations of the present
method.  An approach that can properly deal with low bias situations is still
in development.

Our configuration of the qubit-QPC coupling scheme can be straightforwardly
extended to $N$($N>2)$-qubit detection.  However, it depends strongly on the
sensitivity of the current readout circuit such that the $2^N$ states can be
differentiated,\cite{Tanamoto} and is thus better suited for only a small
number of qubits.  In any case, the key objective of the present study is to
obtain two-qubit information directly and dynamically, not to invent a
general detector for a multi-qubit system, for which other configurations
such as a typical one-detector-per-qubit setup are probably more suitable and
have to be further studied both experimentally and
theoretically.\cite{Gardelis,Cain,Potok,Elzerman,DiCarlo} Furthermore, we
have considered an ideal measurement process in the present study.  In a
more realistic situation, imperfections such as gate operation
errors,\cite{Hu} charge fluctuations around the qubit-QPC
systems,\cite{Itakura} and phonons have to be considered.  These
imperfections could seriously reduce the sensitivity of a measuring device. 
Thus more detailed analysis for the coupled multiqubit-detector system needs
to be carried out in the future to further clarify these issues.

%%%%%%%%%%%%%%%%%%%%%%%%%%%%%%%%%%%%%%%%%%%%%%
\section{Conclusion}

We have solved master equations and described various time-dependent
measurement processes of two charge qubits by two serially-coupled QPCs.
The current through the QPCs is shown to be an effective means to measure the
results of quantum calculations and the presence of entangled states.  
Two-qubit dynamics is studied analytically and it is found that period of 
coherent oscillation depends on their initial state. Our
results thus show that the serially-coupled QPC can be an effective detector
of two-qubit states of a pair of (coupled) charge qubits.

%%%%%%%%%%%%%%%%%%%%%%%%%%%%%%%%%%%%%%%%%%%%%%
\section*{Acknowledgements}
We thank N. Fukushima, S. Fujita, M. Ueda, T. Fujisawa
and S. Ishizaka for valuable discussion.  
Also, XH is grateful to ARO and ARDA of the US for support.

\appendix
\begin{widetext}
\section{Derivation of density matrix equations}
Here we display all the density matrix equations of the qubit-QPC system. 
The density matrix equations can be classified according to the electronic
states of the QPC island (See Fig.~\ref{states}) and qubit states
($z_1,z_2=A,B,C,D$, $s=\uparrow,\downarrow$).
%%%%%%%%%%%%%%%%%%%%%%%%%%%
%%%%%%%%%%%%%%%%%%%   a
%\small
\begin{eqnarray}
%%%(AB)%%%   a
\frac{d \rho^{z_1z_2}_{a}}{dt}\!&=&\!(\!i[J_{z_2}\!-\!J_{z_1}\!]
-\![\Gamma_L^{(z_1)}\!+\!\Gamma_L^{(z_2)} ])\rho^{z_1z_2}_{a}
\!-\! i\Omega_R (\rho^{g_r(z_1),z_2}_{a}\!-\!\rho^{z_1,g_r(z_2)}_{a})
\!-\! i\Omega_L (\rho^{g_l(z_1),z_2}_{a}\!-\!\rho^{z_1,g_l(z_2)}_{a})
\nonumber \\
\!&\!+\!&\! \sqrt{\Gamma_R^{(z_1)}\Gamma_R^{(z_2)}} 
(\rho^{z_1z_2}_{b\uparrow}+\rho^{z_1z_2}_{b\downarrow}),
\\
%%%(AB)%%%   (b)
\frac{d \rho^{z_1z_2}_{b_s}}{dt}\!\!&=&\!\!\left(\!i[J_{z_2}\!-\!J_{z_1}]\!
-\!\frac{\Gamma_L^{(z_1)'} \!+\!\Gamma_L^{(z_2)'} 
\!+\!\Gamma_R^{(z_1)}\!+\!\Gamma_R^{(z_2)}}{2}\right)
\rho^{z_1z_2}_{b_s}
\!-\! i\Omega_R (\rho^{g_r(z_1),z_2}_{b_s}\!-\!\rho^{z_1,g_r(z_2)}_{b_s})
\nonumber \\
\!&\!-\!&\! i\Omega_L
(\rho^{g_l(z_1),z_2}_{b_s}\!-\!\rho^{z_1,g_l(z_2)}_{b_s})
\!+\! 
\sqrt{\Gamma_L^{(z_1)}\Gamma_L^{(z_2)}} \rho^{z_1z_2}_{a}
\!+\! \sqrt{\Gamma_R^{(z_1)'}\Gamma_R^{(z_2)'}} \rho^{z_1z_2}_{c},
\\
%%%(AB)%%%   c
\frac{d \rho^{z_1z_2}_{c}}{dt}
\!\!&=&\!\!(\!i[J_{z_2}\!-\!J_{z_1}]
\!-\![\Gamma_R^{(z_1)'}+\Gamma_R^{(z_2)'}] )
\rho^{z_1z_2}_{c}
\!-\! i\Omega_R (\rho^{g_r(z_1),z_2}_{c}\!-\!\rho^{z_1,g_r(z_2)}_{c})
\!-\! i\Omega_L (\rho^{g_l(z_1),z_2}_{c}\!-\!\rho^{z_1,g_l(z_2)}_{c})
\nonumber \\
\!&\!+\!&\! \sqrt{\Gamma_L^{(z_1)'}\Gamma_L^{(z_2)'}}
(\rho^{z_1z_2}_{b\uparrow}+\rho^{z_1z_2}_{b\downarrow}),
\label{eqn:dm-c}
\end{eqnarray}
\end{widetext}
\normalsize
where
\begin{eqnarray}
\begin{array}{lll}
\Gamma_L^{(A)}&=\Gamma_L^{(B)}=\Gamma_L^{(\!+\!)}, 
&\Gamma_L^{(C)}=\Gamma_L^{(D)}=\Gamma_L^{(\!-\!)}, \nonumber \\
\Gamma_R^{(A)}&=\Gamma_R^{(C)}=\Gamma_R^{(\!+\!)}, 
&\Gamma_R^{(B)}=\Gamma_R^{(D)}=\Gamma_R^{(\!-\!)}, \nonumber \\
\Gamma_L^{(A)'}&=\Gamma_L^{(B)'}=\Gamma_L^{(\!+\!)'}, 
&\Gamma_L^{(C)'}=\Gamma_L^{(D)'}=\Gamma_L^{(\!-\!)'}, \nonumber \\
\Gamma_R^{(A)'}&=\Gamma_R^{(C)'}=\Gamma_R^{(\!+\!)'}, 
&\Gamma_R^{(B)'}=\Gamma_R^{(D)'}=\Gamma_R^{(\!-\!)'},
\end{array} 
\end{eqnarray}
and
\begin{eqnarray}
\begin{array}{lll}
 J_A \!\!&=\! \Delta_L\!+\!\Delta_R\!+\!J, 
&J_B \!\!=\! \Delta_L\!-\!\Delta_R\!-\!J, \nonumber \\ 
 J_C \!\!&=\!-\Delta_L\!+\!\Delta_R\!-\!J, 
&J_D \!\!=\!-\Delta_L\!-\!\Delta_R\!+\!J.
\end{array} 
\end{eqnarray}
$g_l(z_i)$ and $g_r(z_i)$ are introduced for the sake of notational 
convenience and represent relationships between different two-qubit states in
the equations for the density matrix elements:
%%%%%%
\begin{eqnarray}
\begin{array}{lll}
g_r(A)&=B, g_l(A)=C, 
&g_r(B)=A, g_l(B)=D, \nonumber \\
g_r(C)&=D, g_l(C)=A, 
&g_r(D)=C, g_l(D)=B. \nonumber
\end{array}
\end{eqnarray}
%
%The off-diagonal term such as $\sqrt{\Gamma_R^{(\!+\!)}\Gamma_R^{(\!-\!)}}$ 
%is derived from $2\pi\rho_R V_R^{(\!+\!)}V_R^{(\!-\!)}$. 

%%%%%%%%%%%%%%%%%%%%%%%%%%%%%%%%%%%%%%%%%%%%%%


\begin{thebibliography}{99}
\bibitem{footnote} Compared with the two-island SET detector we considered
before,\cite{TanaHu} where we need 352 equations to describe the coupled
qubit-detector system, the number of density matrix equations for the QPC
detector is significantly reduced.
\end{thebibliography}
\end{document}